\documentclass[aps,preprint]{revtex4}

\usepackage{amssymb}
\usepackage{amsmath}
\usepackage{epsfig}
\usepackage{epstopdf}
\usepackage{bm}
\usepackage{graphicx,epsfig}
\usepackage{mathrsfs}
\usepackage{dcolumn}
\usepackage{color}
\usepackage{natbib}
\usepackage{CJK}
\def\be{\begin{equation}}
\def\ee{\end{equation}}
\def\bea{\begin{eqnarray}}
\def\eea{\end{eqnarray}}

\allowdisplaybreaks[2]

\begin{document}

\title{Photonic gauge potential in a system with a synthetic frequency dimension}
\author{Luqi Yuan, Yu Shi, and Shanhui Fan}
\affiliation{Department of Electrical Engineering, and Ginzton
Laboratory, Stanford University, Stanford, CA 94305, USA;}

\date{\today }

\begin{abstract}
We generalize the concept of photonic gauge potential in real
space, by introducing an additional ``synthetic'' frequency
dimension in addition to the real space dimensions. As an
illustration we consider a one-dimensional array of ring
resonators, each supporting a set of resonant modes having a
frequency comb with spacing $\Omega$, undergoing a refractive
index modulation at the modulation frequency equal to $\Omega$. We
show that the modulation phase provides a gauge potential in the
synthetic two-dimensional space with the dimensions being the
frequency and the spatial axes. Such gauge potential can create a
topologically protected one-way edge state in the synthetic space
that is useful for high-efficiency generation of higher-order side
bands.
\end{abstract}

%\pacs{}

\maketitle

The creation of photonic gauge potential in real space opens a new
dimension in the control of light propagation. Such real-space
photonic gauge potential can be created either in time-reversal
invariant systems such as a static resonator lattice or
metamaterials \cite{hafezi11,umucalilar11,fu15}, or in systems
where time-reversal symmetry is broken with either magneto-optical
\cite{fang13pra} or dynamic modulation effects \cite{fang12np}. A
proper choice of photonic gauge potential can lead to an effective
magnetic field for photons \cite{hafezi11,fang12np,hafezi13},
which directly results in the creation of topologically protected
one-way edge states. The ability to specify arbitrary gauge
potential in real space, moreover, can lead to novel effects,
including negative refraction \cite{fang13}, directional-dependent
total internal reflection \cite{fang13}, and gauge-field
waveguides \cite{lin14}. The concept of photonic gauge potential
in real space is also intimately connected to the concept of
photonic gauge potential in momentum space
\cite{onoda04,raghu08,haldane08,wang08,wang09,fang11}, both of
which are of significant importance in the emerging area of
topological photonics
\cite{khanikaev13,rechstman13,mittal14,lu14,yuanFTI}.

\begin{figure}[h]
\includegraphics[width=8cm]{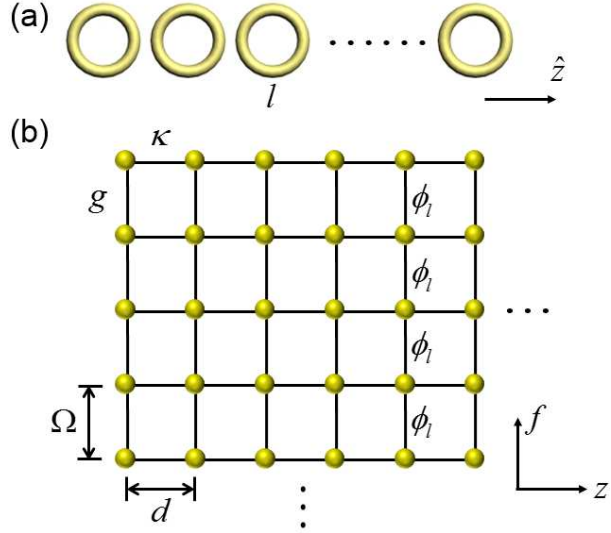}
\caption{(a) A one-dimensional array of ring resonators. Each ring
resonator supports a set of resonant modes with the frequencies of
the modes forming a frequency comb with equally spaced $\Omega$.
The $l$-th ring undergoes a modulation at the modulation frequency
$\Omega$ with a modulation phase $\phi_l$. (b) The system in the
panel (a) can be mapped into a tight-binding model in
two-dimensions, with the extra synthetic dimension being the
frequency dimension. \label{Fig:scheme}}
\end{figure}

In this Letter we generalize the concept of the photonic gauge
potential in real space, by adding a ``synthetic'' frequency
dimension to the real space dimensions. As an illustration, we
start by considering a simple one-dimensional coupled resonator
model shown in Figure \ref{Fig:scheme}(a). Each resonator supports
a set of modes with their frequencies equally spaced at a
frequency $\Omega$, forming a frequency comb. We assume coupling
only between modes having the same frequency at the nearest
neighbor resonators. In addition, we assume that each resonator is
modulated at the frequency $\Omega$, which induces coupling
between modes in the same resonator with frequencies separated by
$\Omega$. The Hamiltonian of the system is then:
\begin{equation}
H= \sum_{l,m} \omega_m a_{l,m}^\dagger a_{l,m}  + \sum_{l,m}
\left\{ \kappa \left( a_{l,m}^\dagger a_{l+1,m} +
a_{l+1,m}^\dagger a_{l,m} \right) + 2g \cos(\Omega t + \phi_l)
\left(a_{l,m}^\dagger a_{l,m+1} + a_{l,m+1}^\dagger a_{l,m}
\right) \right\}, \label{Eq:Hamtotal}
\end{equation}
where $a_{l,m}^\dagger (a_{l,m})$ is the creation (annihilation)
operator for the $m$-th mode at the $l$-th resonator, $\omega_m =
\omega_0 + m\Omega$ gives the frequency for the $m$-th resonant
mode, $\kappa$ is the coupling constant between two
nearest-neighbor resonators, $g$ is the strength of modulation,
and $\phi_l$ is the associated modulation phase at the $l$-th
resonator. Below we will show that this Hamiltonian is equivalent
to a tight-binding Hamiltonian in two dimensions subject to an
out-of-plane effective magnetic field, with each lattice site
corresponding to one mode in a specific resonator, and with the
two dimensions corresponding to the one-dimensional space and the
frequency axes, respectively. In this construction the frequency
axis therefore becomes an extra synthetic dimension.

The concept of synthetic dimension in the study of artificial
gauge field has been recently discussed for cold atoms
\cite{celi14}, where the synthetic dimension corresponds to the
atomic states, and in a photonic system where the synthetic
dimension corresponds to the orbital angular momentum of light
\cite{luo14}. In contrast to these works, here we show that the
use of the frequency space as the synthetic dimension offers new
possibilities for controlling the frequencies of light.

\begin{figure}[h]
\includegraphics[width=12cm]{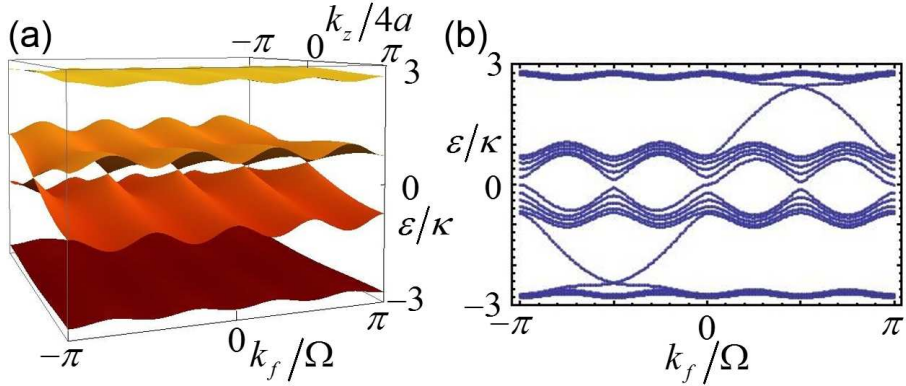}
\caption{(a) Band structure for the system shown in Figure
\ref{Fig:scheme}, as described by the Hamiltonian in Eq.
(\ref{Eq:Hamk}) in the range $-\pi/4a < k_z < \pi/4a$ and
$-\pi/\Omega<k_f<\pi/\Omega$. (b) Projected band structure for the
system consisting of 21 resonators. \label{Fig:bands}}
\end{figure}

To see the emergence of an effective magnetic field in Eq.
(\ref{Eq:Hamtotal}), we define $c_{j,m} \equiv a_{j,m}e^{-i
\omega_m t}$, and apply the rotating wave approximation to obtain
from Eq. (\ref{Eq:Hamtotal})
\begin{equation}
H= \sum_{l,m} \left\{ \kappa \left( c_{l,m}^\dagger c_{l+1,m} +
c_{l+1,m}^\dagger c_{l,m} \right) + g \left( e^{-i\phi_l}
c_{l,m}^\dagger c_{l,m+1} + e^{i\phi_l} c_{l,m+1}^\dagger c_{l,m}
\right) \right\}. \label{Eq:Ham}
\end{equation}
Eq. (\ref{Eq:Ham}) is identical in form  to the Hamiltonian of a
quantum particle on a two dimensional lattice subject to a
magnetic field \cite{fang12np}, except that here, one of the axes
(as labelled by the modal index $m$) is the frequency axis. Since
there is one modulator per resonator, the hopping phases are
uniform along the frequency axis. Nevertheless, Eq. (\ref{Eq:Ham})
is sufficient to implement the Landau gauge for a uniform magnetic
field by choosing
\begin{equation}
\phi_l = l \cdot \pi/2. \label{Eq:phasedist}
\end{equation}
The resulting system has the periodicity of $4d$ along the
$z$-direction, where $d$ is the physical spacing between the
nearest neighbor resonators, and $\Omega$ along the frequency
dimension. For an infinite lattice, we can re-write Eq.
(\ref{Eq:Ham}) into the $\bm{k}$-space
\begin{equation}
H_k= \sum_{\alpha=1}^4 \left\{ \kappa \left( a_{k,\alpha}^\dagger
a_{k,\alpha+1} e^{i k_z d} + a_{k,\alpha+1}^\dagger a_{k,\alpha}
e^{-i k_z d} \right) +2g  a_{k,\alpha}^\dagger a_{k,\alpha}
\cos\left(k_f \Omega - \phi_\alpha\right) \right\},
\label{Eq:Hamk}
\end{equation}
where $\phi_\alpha = \pi/2,\pi,3\pi/2,2\pi$, respectively. We plot
the band structure, as defined as the eigenvalues of the
Hamiltonian in (\ref{Eq:Hamk}) as a function of $\bm{k}$ in Figure
\ref{Fig:bands}(a) with $g=\kappa$. There are 4 bands with Chern
number -1, 1, 1, -1, respectively. The band gaps between the top
two bands  and between the bottom two bands are therefore
topologically non-trivial.

For a finite lattice, the non-trivial topology manifests in the
creation of topologically protected one-way edge states in the
eigenvalue ranges of the band gaps. As an illustration, we
consider a system as described by the Hamiltonian of Eq.
(\ref{Eq:Ham}) with 21 resonators, assuming that the frequency
axis is still infinite. The projected bandstructure, defined as
the eigenvalues of Eq. (\ref{Eq:Ham}) with respect to the
wavevector $k_f$ along the frequency axis, is shown in Figure
\ref{Fig:bands}. We indeed observe two one-way edge states in the
band gaps. Unlike the standard tight-binding models with two
spatial dimensions, however, here the edge states corresponds to
propagation in the \textit{frequency} space. As a result the
concept of a synthetic dimension of frequencies allows novel
possibility for manipulating the frequency of light.

We now provide a physical system that implements the idea of
synthetic dimension as described in Eq. (\ref{Eq:Hamtotal}). For
this purpose we consider a one-dimensional array of optical ring
resonators. Each ring resonator consists of a single-mode
waveguide forming a loop. The waveguide has a dispersion relation:
\begin{equation}
\beta = n(\omega)\frac{\omega}{c}, \label{Eq2:dispersion}
\end{equation}
where $\beta$ is the wavevector along the propagation direction,
which will be denoted as the $x$-direction. $n(\omega)$ is the
effective phase index. For a ring with a circumference $L$,
suppose it supports a resonance at $\omega_0$, i.e.
$\beta(\omega_0) L  = 2 \pi m_0$. In the vicinity of $\omega_0$,
the resonance condition is
\begin{equation}
[\beta(\omega_m')-\beta(\omega_0)]L = 2\pi m,
\label{Eq2:condition}
\end{equation}
where $\omega_m'$ is the resonant frequency of the $m$-th order
mode. If one ignores group velocity dispersion, then
\begin{equation}
\omega_m' = \omega_m  \equiv \omega_0 + m  \Omega,
\end{equation}
where $\Omega \equiv 2 \pi c/Ln_g(\omega_0)$, with the group index
$n_g(\omega_0)= n + \omega \frac{\partial n}{\partial
\omega}|_{\omega_0}$, and $m$ is an integer. Below we refer to the
fields at $\omega_m$ as the $m$-th side band. On the other hand,
with group velocity dispersion, i.e. when $\partial n_g /
\partial \omega \neq 0$,  $\omega_m'$ and $\omega_m$ are no longer
the same.

Suppose we modulate the ring at the modulation frequency $\Omega$.
The electric field inside the ring in general can be written as
\begin{equation}
E (t,r_\bot,x)= \sum_m \mathcal{E}_m(t,x)E_m(r_\bot)e^{i\omega_m
t},
\end{equation}
where $r_\bot$ denotes the directions perpendicular to the
waveguide, $E_m(r_\bot)$ is the modal profile of the waveguide.
$\mathcal{E}_m(t,x)$ is the modal amplitude associated with the
$m$-th sideband. Assuming that the modal amplitudes are slowly
varying, we then have \cite{hausbook}
\begin{equation}
\left(\frac{\partial}{\partial x} + i\beta (\omega_m) \right)
\mathcal{E}_m - \frac{n_g(\omega_m)}{c} \frac{\partial}{\partial
t}\mathcal{E}_m = 0. \label{Eq2:maxsol}
\end{equation}
with $\mathcal{E}_m(t,x+L) = \mathcal{E}_m(t,x)$.

\begin{figure}[!h]
\includegraphics[width=10cm]{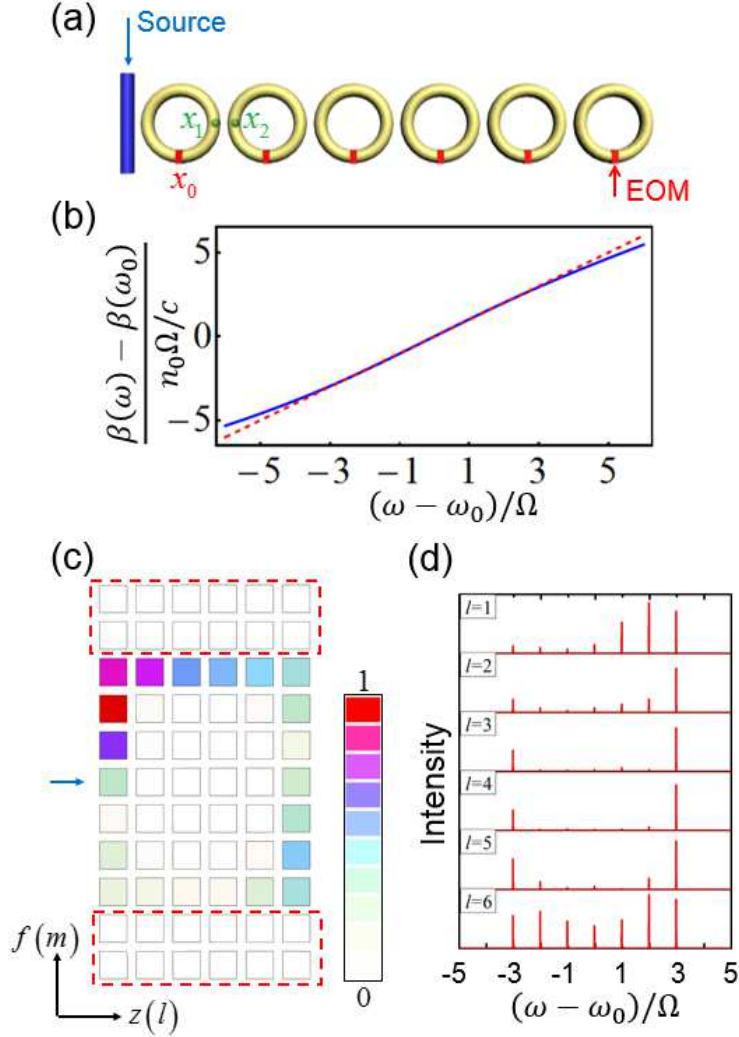}
\caption{(a) An array of six ring resonators. The red dots
represent the electro-optic modulator (EOM) on the resonators. The
green dots represent the positions where coupling occur between
nearest neighbor resonators. An input waveguide couples to the
left-most ring resonator. (b) The dispersion relation of the
waveguide that forms the ring. (c) The distribution of the total
intensity in each ring resonator for each frequency component
($|\mathcal{E}_{l,m}|^2$). Blue arrow indicates the continuous
source. The frequency components in the dashed red box corresponds
to the non-resonant frequency sidebands. (d) The intensity spectra
for the field $E(t,r)$ at each resonator. \label{Fig:sim1}}
\end{figure}

We describe the coupling of the fields between the $l$-th and
$(l+1)$-th resonators as:
\begin{equation}
\mathcal{E}_m^l (t^+,x_1^l)  = \sqrt{1-\gamma^2} \mathcal{E}_m^l
(t^-,x_1^l)-i\gamma \mathcal{E}_m^{l+1} (t^-,x_2^{l+1}),
\label{Eq2:coupledmode1}
\end{equation}
\begin{equation}
\mathcal{E}_m^{l+1} (t^+,x_2^{l+1})  = \sqrt{1-\gamma^2}
\mathcal{E}_m^{l+1} (t^-,x_2^{l+1})-i\gamma \mathcal{E}_m^{l}
(t^-,x_1^{l}), \label{Eq2:coupledmode2}
\end{equation}
where $\gamma$ represents the coupling strength. $x_1^{l}$ and
$x_2^{l+1}$ are the positions on the two rings where the coupling
occur (see Figure \ref{Fig:sim1}(a)), and $t^\pm = t + 0^\pm$.

We assume that the dynamic modulation of the $l$-th ring is
achieved by placing a phase modulator at $x_0^l$ (Figure
\ref{Fig:sim1}(a)). Upon passing through $x_0^l$, the total
electric field undergoes a phase modulation $E \rightarrow
Ee^{i\alpha\sin(\Omega t + \phi)}$ \cite{Saleh}, or equivalently
\begin{equation}
\mathcal{E}_m^l (t^+,x_0^l) = J_0 (\alpha) \mathcal{E}_m^l
(t^-,x_0^l) + \sum_{q} J_q (\alpha) \mathcal{E}_{m-q}^l
(t^-,x_0^l) e^{iq\phi} + \sum_{q} (-1)^q J_q (\alpha)
\mathcal{E}_{m+q}^l (t^-,x_0^l) e^{-iq\phi},
\label{Eq2:phasemodulation}
\end{equation}
where $J_q$ is the Bessel function of the $q$-th order. The
modulation therefore induces a coupling between sidebands, with
the modulation phase appearing as the phase in the coupling
constant, similar to the tight-binding model of Eqs.
(\ref{Eq:Hamtotal}) and (\ref{Eq:Ham}). Unlike the tight-binding
model, here there is a long-range coupling in the frequency space.
We will show in the simulation below that such a long range
coupling does not fundamentally affect the physics of synthetic
dimension as described simply in the tight-binding model.

The excitation is injected into the system through an input/output
waveguide:
\begin{equation}
\mathcal{E}_m^l (t^+,x_e^l)  = \sqrt{1-\eta^2} \mathcal{E}_m^l
(t^-,x_e^l)-i\eta \mathcal{E}_s (t^-,x_e^l), \label{Eq2:output}
\end{equation}
where $\eta$ is the strength of the coupling between the resonator
and the waveguide and $\mathcal{E}_s$ is the source field in the
waveguide.

Eqs. (\ref{Eq2:maxsol}-\ref{Eq2:output}) provide a description of
a physical system consisting of a set of ring resonators coupled
together, with each ring modulated by an electro-optic phase
modulator. We solve Eq. (\ref{Eq2:maxsol}) with a finite
difference approach in both space and time.  At the places where
the coupling occurs or at the locations of the modulators, we
calculate the fields at time $t^-$ first with Eq.
(\ref{Eq2:maxsol}) and then apply either Eqs.
(\ref{Eq2:coupledmode1}) and (\ref{Eq2:coupledmode2}) or Eq.
(\ref{Eq2:phasemodulation}) to compute the fields at $t^+$.

In the modulated ring, with the absence of group velocity
dispersion, the modulation sidebands at $\omega_m$ coincide with
the modes of the ring at $\omega_m'$. Therefore we have
on-resonance coupling between multiple modes. On the other hand,
with group velocity dispersion, $\omega_m' \neq \omega_m$, and the
coupling between the modes become off-resonance. As a result, the
group velocity dispersion of a waveguide provides a natural
``boundary'' in the frequency space. As an illustration, we first
perform a simulation of a system with six ring resonators as shown
in Figure \ref{Fig:sim1}(a). We assume that at a center frequency
$\omega_0$ the waveguide for the ring has an effective index of
$n_0 = n(\omega_0)$=1.5. In the simulation, we include 11 side
bands ($\omega_m=\omega_0+m\Omega$ and $m=-5,-4,\ldots,5$), where
$\Omega = 2 \pi c/n_0 L$. We choose a waveguide dispersion
relation $\beta(\omega)$, as shown in Figure \ref{Fig:sim1}(b),
such that 7 of the side bands with $m=-3,-2,\ldots,3$ are
on-resonance having $n(\omega_m) = n_0$. The other 4 side bands
have an effective index that differs from $n_0$. The dispersion
relation here thus is chosen to illustrate a waveguide with a zero
group velocity dispersion near frequency $\omega_0$. Each of the
rings is modulated as described above with an electro-optic
modulator with a modulation frequency $\Omega$, and with a
modulation phase $\phi_l$ in Eq. (\ref{Eq:phasedist}). To excite
the system, a continuous-wave signal, having a single frequency at
$\omega_{m=0}$, is sent into the left ring resonator ($l=1$). The
distribution of the intensity $|\mathcal{E}_{l,m}|^2$ as a
function of $l$ and $m$ at $t=400$ $n_0L/c$ (when the field
finishes one loop in the synthetic space) is plotted in Figure
\ref{Fig:sim1}(c). We note that there is almost zero intensity for
the sidebands with $m = \pm 5$ and $\pm 4$. Thus the group
velocity dispersion indeed provides a boundary in the frequency
space. The intensity is concentrated at the edge of the synthetic
space forming a topologically protected one-way mode. We plot in
Fig. \ref{Fig:sim1}(d) the intensity spectra corresponding to the
$E(t,r)$ field inside each resonators. For the resonators at the
spatial edge, ($l = 1$, and $l = 6$), the intensity spectra have
significant components in all on-resonance side bands, while for
the resonators at the center of the structure ($l = 3$, and $l
=4$), the intensity spectra are almost completely concentrated in
the on-resonance side bands that have the highest and lowest
frequencies.

\begin{figure}[h]
\includegraphics[width=8cm]{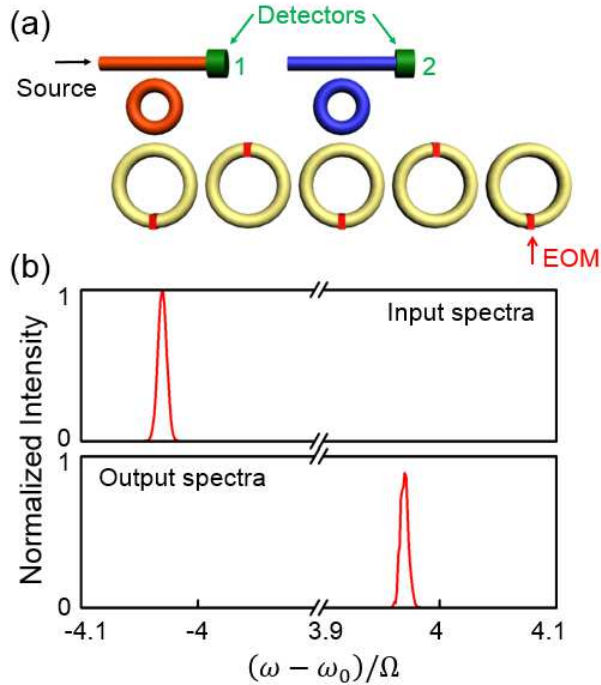}
\caption{(a) An array of five ring resonators. Two extra
single-mode resonators couple between the array and the waveguides
with detectors. (b) The intensity spectra of the input and output.
The output spectra is measured at detector 2. \label{Fig:sim2}}
\end{figure}

In a typical ring-resonator system under modulation, multiple side
bands are generated. On the other hand, in the system considered
here the resonators in the center of the structure have amplitudes
only in the highest and lowest on-resonance frequency side bands.
Therefore, the use of a gauge field in the synthetic space
provides the opportunity to create high-efficiency generation of
high-order side-bands. As an illustration, we consider the system
shown in Figure \ref{Fig:sim2}(a). The system consists of five
resonators modulated at a frequency $\Omega$ with the modulation
phase chosen according to Eq. (\ref{Eq:phasedist}) such that we
have a non-zero gauge field in the synthetic space. In the
simulations, we include 11 side bands ($m = -5, \ldots +5$).  We
choose the waveguide dispersion relation such that the side bands
at $m = \pm 5$ are non-resonant and the rest of the side bands are
resonant with the ring modes. We couple to the ring resonator at
$l = 1$ ($l = 3$) with an extra ring resonator having a resonance
at $\omega_{m=-4}$, ($\omega_{m=+4}$) and are off resonance at all
other side-band frequencies. To excite the system, we inject into
the additional resonator associated with the $l=1$ resonator with
a pulse which has its carrier frequency $\omega_{m=-4}$ and the
temporal Full Width at Half Maximum (FWHM) $\Delta t = 50$
$n_0L/c$, corresponding to the FWHM of the spectral intensity
$\Delta \omega = 0.055$ $c/n_0 L \ll \Omega$. Therefore the input
pulse only excites a single side band. The topologically protected
one-way edge state converts the input at $\omega_{m=-4}$ to the
frequency component at $\omega_{m=4}$ when the signal propagates
along the edge of the lattice. We plot the input and output
intensity spectra and the convert spectra in Figure
\ref{Fig:sim2}(b). The input field has a single frequency near
$\omega_{m=-4}$ to excite the edge mode in the band gap. The
output spectra is measured at detector 2 and we see that the field
has been converted to the field with the frequency near
$\omega_{m=4}$ with an efficiency of $81\%$. Our result shows that
this system serves the purpose of a higher order frequency
converter.

Our proposal here can be implemented in various systems in fiber
optics or integrated photonics. For the standard fiber system, the
electro-optic phase modulation frequency can be up to $\sim 1$
GHz. This requires the resonator composed by the loop of a fiber
with the length of $\sim 0.1$ m. We note that the standard
single-mode fiber has a zero group velocity dispersion point at
$\lambda = 1.3$ $\mathrm{\mu}$m, which is useful for our design
here. For a silicon resonator modulated at $\sim 100$ GHz, the
radius of the resonator is $\sim 100$ $\mathrm{\mu}$m
\cite{tzuang14}. A flat and low in-cavity dispersion over a wide
wavelength range ($\sim 500$ nm) can be achieved in a curving
waveguide \cite{zhang11}. Further miniaturization of the structure
in either platform can be accomplished with the use of slow-light
waveguides.

In summary, we generalize the concept of photonic gauge potential
to a synthetic space with both the spatial and frequency
dimensions, and demonstrate that such a gauge potential leads to
an edge state in the synthetic space that is useful for
high-efficiency conversion to higher-order side bands. Related to
and independent of our work, a recent preprint \cite{ozawa15} has
proposed the use of similar synthetic gauge potential for
simulation of quantum hall effect in four-dimensions. We believe
that the concept of gauge potential in such synthetic space
provides a new dimension for the control of light in both the real
and the frequency spaces.

\begin{acknowledgments}
This work is supported in part by U.S. Air Force Office of
Scientific Research Grant No. FA9550-12-1-0488. We also
acknowledge discussions with I. Carussotto, who alerts us to Ref.
\cite{ozawa15} when we were in the final stage of preparing this
manuscript.
\end{acknowledgments}

\end{document}